# Spray flame synthesis of $Y_2O_3$-MgO nanoparticles for mid-infrared transparent nanocomposite ceramics

Shuting Lei[a], Yiyang Zhang[b], Xing Jin[c], Yanan Li[c], Zhu Fang[a], Shuiqing Li[a]

[a]Key Laboratory for Thermal Science and Power Engineering of Ministry of Education, Department of Energy and Power Engineering, Tsinghua University, Beijing, 100084, China

[b]Key Laboratory of Advanced Reactor Engineering and Safety of Ministry of Education, Collaborative Innovation Center of Advanced Nuclear Energy Technology, Institute of Nuclear and New Energy Technology, Tsinghua University, Beijing 100084, China

[c]Research Center for Gas-phase Synthesis of Functional Nanomaterials, Wuzhen Laboratory, Jiaxing 314500, China

**Abstract:** Spray flame synthesis offers a promising method for scalable production of homogeneously mixed $Y_2O_3$–MgO nanopowders as next-generation infrared-transparent window material, which has attracted significant attention owing to its excellent optical properties at high temperatures. However, systematic understanding of how flame synthesis parameters influence particle morphology, crystal phase, solid solubility, and subsequent ceramic performance remains insufficiently understood. In this study, we investigated the influence of precursor chemistry on particle crystal phase and examined the solid solubility of MgO in $Y_2O_3$ under different flame temperatures, demonstrating that the high-temperature conditions with $O_2$ as dispersion gas allow up to 50 mol% MgO to fully dissolve into $Y_2O_3$, far exceeding the equilibrium solubility limit of ~7 mol% at the eutectic temperature (2100°C) and near-zero at room temperature. Furthermore, we systematically evaluated how powder characteristics and sintering parameters—including powder deagglomeration methods, vacuum sintering temperature, hot isostatic pressing (HIP) temperature, and initial powder characteristics—affect ceramic microstructures and infrared transmittance. Despite cracking induced by phase transformation and finer particle sizes, ceramics fabricated from oxygen-synthesized monoclinic-dominated powders exhibited superior near-infrared transmittance (56.2% at 1550 nm), attributed to enhanced atomic mixing and effective grain boundary pinning. After optimization, pure cubic phase powders produced intact and crack-free ceramics with outstanding mid-infrared transparency, achieving a maximum transmittance of 84.6% and an average transmittance of 82.3% in 3–5 μm range.

## 1. Introduction

As the foremost component of an aircraft's imaging system, the infrared (IR) transparent window plays a critical role in determining the performance of target detection, guidance, and recognition[1-3]. With the increasing demands of high-speed flight, especially in hypersonic regimes, window materials should be able to endure extreme environments including extreme thermal loads, high-velocity airflow, mechanical shock, and chemical corrosion. Therefore, high performance window should be able to maintain high IR transmittance and low emittance at elevated temperatures to avoid self-emission that interferes with imaging, as well as excellent mechanical strength, thermal shock resistance, and chemical stability[3-5]. Conventional materials like zinc sulfide (ZnS), once widely used in IR applications, lack the mechanical robustness required for high-speed flight. Sapphire, despite offering excellent room-temperature performance, suffer from a significant increase in emittance and degradation in strength at high temperatures[6]. In recent years, yttria ($Y_2O_3$) has gained attention due to its low emittance and stability under elevated temperatures. However, conventional processing of polycrystalline $Y_2O_3$ often results in significant grain growth due to the prolonged high-temperature sintering, leading to inferior mechanical properties[7]. To suppress grain coarsening, one effective strategy is to introduce a second immiscible phase, which hinders grain boundary migration during sintering—a phenomenon commonly referred to as "pinning effect"[8]. Magnesia (MgO), which exhibits excellent IR transparency and minimal mutual solubility with $Y_2O_3$ at room temperature (see the $Y_2O_3$–MgO phase diagram[9] in Fig. 1), has been demonstrated to effectively restrain $Y_2O_3$ grain growth and enhance mechanical properties[10, 11]. Consequently, $Y_2O_3$–MgO composite ceramics have emerged as one of the most promising candidates for next-generation IR window materials.

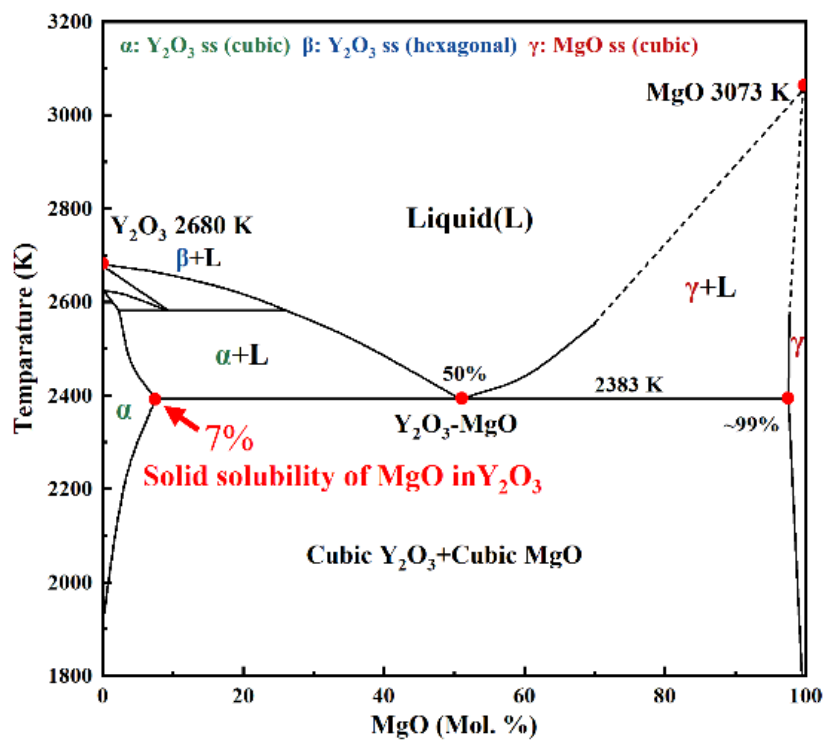

Fig. 1 The phase diagram of $Y_2O_3$-MgO[9]

The key challenge in fabricating IR-transparent composite ceramics lies in the significant difference in refractive indices between the two phases (MgO ~ 1.736, $Y_2O_3$ ~ 1.926 at 600 nm), which causes severe grain boundary scattering [12]. To mitigate this effect, it is essential to minimize the grain size, ideally below 1/15 of the target transmission wavelength[3]. In addition, mechanical performance can be improved by grain refinement, as described by the Hall–Petch relationship[13]. Therefore, to prepare IR transparent composite ceramics with excellent optical and mechanical properties, it is first necessary to achieve the synthesis of $Y_2O_3$–MgO powders with nano-sized, homogeneous mixed elements, and high purity to ensure ultrafine grain size, homogeneous phase distribution in the final composite ceramics.

Although various chemical methods such as co-precipitation[14, 15], sol–gel[16-18], and self-propagating high-temperature synthesis[19, 20] have been employed to synthesize $Y_2O_3$–MgO powders, these methods often involve complex multistep procedures and provide limited control over homogeneity and purity, making large-scale production difficult. In contrast, spray flame synthesis offers a one-step, scalable route capable of producing nano-sized particles with high purity and homogeneous mixing at the atomic scale[21]. Owing to the high flame temperatures and rapid quenching rates, the synthesized powders often exceed the solid solution limits in equilibrium phase diagram[22, 23]. These highly homogeneous, atomically mixed solid solutions can significantly enhance atomic diffusion, promote densification during sintering[24] and facilitates the formation of a uniformly distributed two-phase microstructure in the ceramics. Harris et al.[8] previously demonstrated that fully dense, IR-transparent $Y_2O_3$–MgO ceramics could be fabricated from flame-synthesized powders via vacuum sintering. However, existing research did not systematically address several crucial questions: how the flame synthesis conditions influence the initial powder characteristics; how these initial characteristics subsequently affect the final ceramic performance; and how to identify optimal sintering conditions tailored specifically for flame-synthesized powders.

In this study, we systematically address these gaps by spray flame synthesis. We first investigate the influence of precursor chemistry on particle morphology and crystal phase. Next, we explore the effect of flame temperature on solid solubility. Finally, we evaluate how powder pre-treatment, vacuum sintering temperature, hot isostatic pressing temperature, and initial powder characteristics affect ceramic microstructure and infrared transmittance.

## 2. Experimental methods

*2.1 Flame synthesis of $Y_2O_3$–MgO nanopowders*

The swirl-stabilized spray flame synthesis system used in this study has been described in detail previously[25, 26]. Briefly, methane (3.2 L/min) and air (32 L/min) were introduced tangentially into the burner through four alternating inlet channels, creating a stable, high-speed swirling flame. Precursor solutions, delivered at 600 ml/h by a syringe pump, were atomized into fine droplets at the burner center by a dispersion gas flow (15 L/min). Upon combustion, nanoparticles were carried upward by the gas flow generated by a vacuum pump and collected on a filter membrane. The precursor solutions were prepared by dissolving yttrium nitrate hexahydrate (YN, $Y(NO_3)_3 \cdot 6H_2O$, 99.999%, Mreda) or yttrium acetate tetrahydrate (YCA, $Y(CH_3COO)_3 \cdot 4H_2O$, 99.9%, Adamas), and magnesium nitrate hexahydrate (MgN, $Mg(NO_3)_2 \cdot 6H_2O$, AR, Tongguang) or magnesium acetate tetrahydrate (MgCA, $Mg(CH_3COO)_2 \cdot 4H_2O$, 99%, Aladdin) in anhydrous ethanol (EtOH, ≥ 99.7%) at a metal-ion concentration of 0.6 mol/L. An equivalent amount of 2-ethylhexanoic acid (EHA, > 99.0%, Aladdin) was added to facilitate nanoparticle synthesis. The detailed precursor formulations and flame synthesis conditions are summarized in Table 1. The molar ratio of 20%:80% corresponds to a volume ratio of 50%:50%, which has been proven to be the ratio that exhibits the most effective pinning effect[27].

Table 1 Summary of flame synthesis conditions

| Objective | Precursor of $Y_2O_3$ | Precursor of MgO | Molar ratio of $Y_2O_3$ to MgO | Precursor feed rate | Dispersion gas |
|---|---|---|---|---|---|
| Effects of precursor chemistry | YN<br>YN<br>YCA<br>YCA | MgN<br>MgCA<br>MgN<br>MgCA | 20%:80% | 600 ml/h | 15 L/min Air |
| Solid solubility of MgO in $Y_2O_3$ | YN | MgN | 0%~100% | 600 ml/h | 15 L/min Air<br>15 L/min $O_2$ |

*2.2 Fabrication of $Y_2O_3$-MgO infrared transmittance ceramics*

To fabricate dense $Y_2O_3$–MgO ceramics, a sintering route combining vacuum sintering and hot isostatic pressing (HIP) was adopted. Compared to spark plasma sintering (SPS), vacuum sintering offers lower carbon contamination, reduced processing cost, and greater scalability for producing large or complex-shaped ceramic components.

The as-synthesized powders were first calcined in air at 500 °C for 1 h to remove residual organics and moisture. For studies on pre-treatment and sintering parameters, a secondary calcination step at 1200 °C for 1 h in air was applied to transform the powders into a fully cubic phase. Two powder deagglomeration methods were studied: (1) ball milling in anhydrous ethanol for 20 h; and (2) high-shear homogenization using a dispersing homogenizer. Unless otherwise noted, ball-milled powders were used throughout this work. The deagglomerated powders were dried, sieved, and then uniaxially pressed into cylindrical pellets (20 mm diameter) at 40 MPa, followed by cold isostatic pressing (CIP) at 200 MPa for 5 min. The green bodies were sintered in vacuum for 5 h at designated temperatures. A heating rate of 5 °C/min was used up to 1100 °C, and 3 °C/min beyond that temperature. Then the samples underwent HIP treatment at specified temperatures under 200 MPa of argon for 2 h, using a heating rate of 5 °C/min. All ceramics were subsequently annealed in air at 1000 °C for 20 h to relieve internal stress and reduce oxygen vacancies. Finally, the ceramic pellets were double-side polished to a thickness of 1 mm for optical and structural characterization. The sintering conditions for fabricating the composite ceramics are summarized in Table 2.

Table 2 Summary of conditions for fabricating the composite ceramics

| Objective | Sample ID | Initial powder | Annealed at 1200 °C | Deagglomeration method | Vacuum sintering temperature (°C) | HIP temperature (°C) |
|---|---|---|---|---|---|---|
| Effect of powder deagglomeration | None<br>Ball milling<br>Homogenization | YN-MgN (Air) | Yes | None<br>Ball milling<br>Homogenization | 1400 | 1350 |
| Effect of vacuum sintering temperature | T1400<br>T1350<br>T1300<br>T1250 | YN-MgN (Air) | Yes | Ball milling | 1400<br>1350<br>1300<br>1250 | None |
| Effect of HIP temperature | T1250H1200<br>T1250H1250<br>T1250H1300 | YN-MgN (Air) | Yes | Ball milling | 1250 | 1200<br>1250<br>1300 |
| | T1300H1200<br>T1300H1250<br>T1300H1300 | | | | 1300 | 1200<br>1250<br>1300 |
| Effect of initial powder | 71%Cubic | YN-MgN (Air) | No | Ball milling | 1300 | 1300 |
| | 95% Monoclinic | YN-MgN ($O_2$) | No | | | |
| | 100%Cubic-Annealing | YN-MgN ($O_2$) | Yes | | | |

*2.3 Material Characterization*

The morphology of the as-synthesized particles was examined using transmission electron microscopy (TEM; JEM-F200, JEOL, Japan) operated at 120 kV. Elemental distribution was analyzed using energy-dispersive X-ray spectroscopy (EDS) integrated with the TEM. The crystalline phase of the powders was identified by X-ray diffraction (XRD; D8 Advance, Bruker, Germany) with a scan rate of 2°/min. The specific surface area of the $Y_2O_3$–MgO nanopowders was measured by nitrogen adsorption using the Brunauer–Emmett–Teller (BET) method (IQ2, Quantachrome Instruments, USA). Differential scanning calorimetry (DSC; TGA/DSC1/1600LF, Mettler-Toledo, Switzerland) was used to analyze the thermal behavior of the precursor solutions. The measurements were carried out in air from 50 to 800 °C at a heating rate of 10 °C/min. The bulk density of the sintered ceramics was measured by the Archimedes method in deionized water. The microstructure and phase distribution were observed using scanning electron microscopy (SEM; Gemini 500, Zeiss, Germany). Prior to observation, the polished ceramic surfaces were thermally etched at 1000 °C for 1 h. Grain sizes were measured using the line-intercept method, with at least 300 grains counted. IR transmittance (2.5–10 μm) was measured in transmission mode using Fourier-transform infrared spectroscopy (FT-IR; Nicolet 6700, Thermo Nicolet, USA). Near-infrared to visible transmittance (900–2000 nm) was obtained using a UV–Vis–NIR spectrophotometer (Lambda 950, PerkinElmer, USA).

## 3. Result and discussion

*3.1 Effect of precursor chemistry*

The precursor chemistry has been widely recognized as a critical factor influencing particle morphology, crystal phase, and composition in flame synthesis processes [28, 29]. In this section, four precursor combinations (YN-MgN, YN-MgCA, YCA-MgN and YCA-MgCA) were examined. The effect of precursor combinations on the crystal phase are shown in Fig.

2. All samples exhibited diffraction peaks at 2θ = 42.9° and 62.3°, corresponding to the (200) and (220) planes of cubic MgO (PDF#45-0946), indicating the successful formation of crystalline MgO. Distinct differences were observed in the $Y_2O_3$ phase. The YN-MgN sample exhibited diffraction peaks at 2θ = 16.7°, 29.2°, 33.8°, and 48.5°, which are consistent with the cubic phase of $Y_2O_3$ (C-$Y_2O_3$, PDF#71-0099), suggesting the crystallization into the thermodynamically stable cubic phase. In comparison, the other three combinations containing acetate precursors (YCA-MgCA, YCA-MgN, YN-MgCA) showed additional peaks at 2θ = 28.8°, 29.8°, 30.5°, and 61.7°, which are characteristic of monoclinic $Y_2O_3$ (M-$Y_2O_3$, PDF#44-0399), a metastable phase typically observed under high-temperature or rapid-quench conditions[30]. Among them, the YCA-MgCA sample exhibited the highest intensity of monoclinic phase peaks, suggesting a stronger tendency toward metastable phase formation.

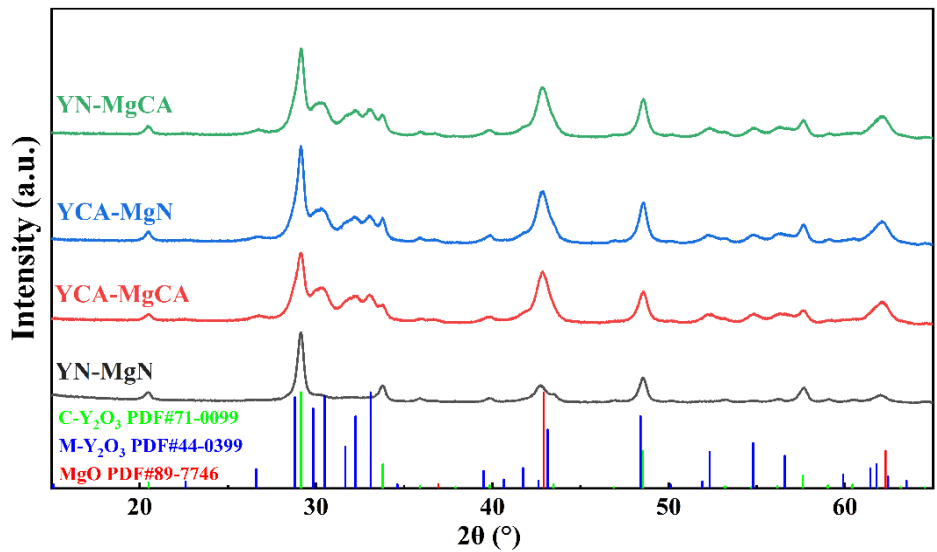

Fig. 2 XRD patterns of four samples with different precursor combinations

Fig. 3(a) shows the crystalline phase composition of the as-synthesized particles with different precursor combinations, quantified by Rietveld refinement. The MgO weight fractions obtained from the refinement are 42%, 45%, 44%, and 43% for YN-MgN, YCA-MgN, YN-MgCA, and YCA-MgCA, respectively, closely matching the theoretical value (42%) determined by the stoichiometry of the precursor solutions. This result highlights the accuracy of flame synthesis in achieving the target phase ratios. In terms of $Y_2O_3$ crystalline phases, YN-MgN exhibits the highest fraction of cubic $Y_2O_3$ (C-$Y_2O_3$) at 41 wt%, accounting for 71% of the total $Y_2O_3$ content. In contrast, the YCA-MgCA sample exhibits the lowest C-$Y_2O_3$ content (13 wt%, 24% of total $Y_2O_3$). The other two combinations, YCA-MgN and YN-MgCA, also show a predominance of the monoclinic $Y_2O_3$ (M-$Y_2O_3$) phase, with the cubic phase comprising only 28% and 32% of total $Y_2O_3$, respectively.

The variations in phase composition are closely related to the thermal behavior of the precursor solutions, as shown in Fig. 3(b). The YCA-MgCA solution exhibits the most intense and sharply defined exothermic peak, indicative of a more concentrated heat release that leads to higher local flame temperatures and steep cooling rates, both of which have been reported to favor the formation of M-$Y_2O_3$[31]. In contrast, the YN-MgN precursor solution shows a broader, weaker thermal peak, suggesting a more gradual decomposition and longer residence at high temperature, which facilitates the crystallization of C-$Y_2O_3$[32]. The YCA-MgN displays the second-highest and sharpest exothermic peak, which corresponds well with its relatively high monoclinic phase content.

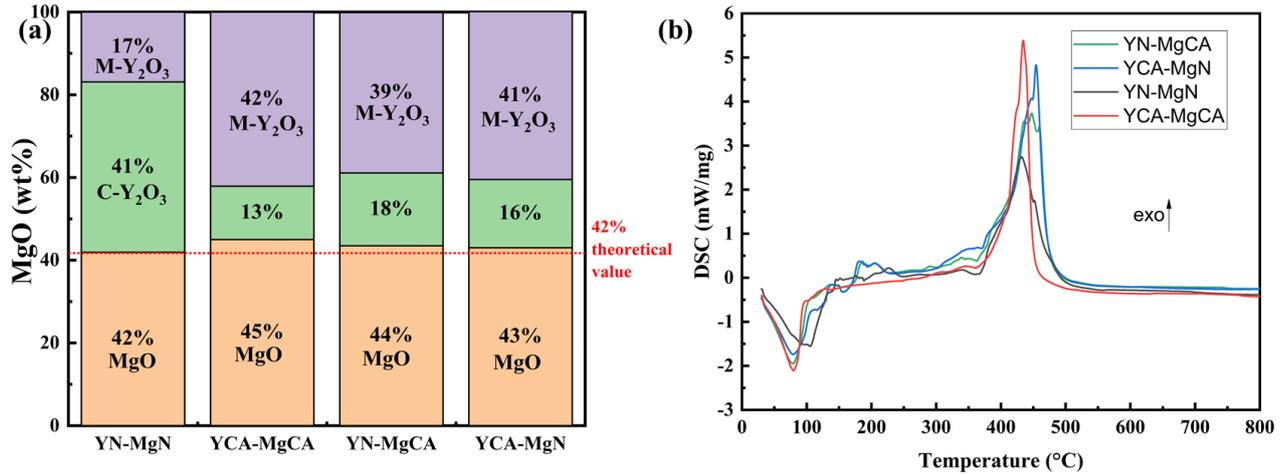

Fig. 3 (a) Phase composition of samples synthesized from different precursor combinations, as quantified by Rietveld refinement of XRD patterns; (b) Differential scanning calorimetry (DSC) curves of the precursor solutions.

Fig. 4(a–d) presents the particle morphologies synthesized from four different precursor combinations. The samples exhibit similar morphologies, consisting of uniformly sized, and nanoscale particles with sintering necks. The corresponding EDS elemental mapping in Fig. 4(e–h) confirms a homogeneous spatial distribution of Y and Mg, suggesting good element mixing during flame synthesis. Table 3 summarizes the particle sizes derived from XRD Rietveld refinement and BET measurements. Across the four precursor combinations, the particle sizes are comparable. C-$Y_2O_3$ exhibits the largest crystallite sizes, ranging from 21.1 to 23.9 nm, while M-$Y_2O_3$ shows significantly smaller sizes (~10 nm). This result is consistent with the flame synthesis results of single $Y_2O_3$, because small particles often cause greater intra-particle stress, thereby forming a monoclinic phase[33]. The crystallite size of MgO remains around 12 nm for all combinations. The BET-derived particle sizes ($d_{BET}$) fall in the range of 18.2–20.8 nm. Among them, YCA-MgCA shows the smallest $d_{BET}$ value, while YN-MgN yields the largest. This trend corresponds well with the DSC results in Fig. 3(b), where a broader, less intense exothermic peak (as seen in YN-MgN) suggests a longer residence time in the high-temperature zone, favoring particle growth.

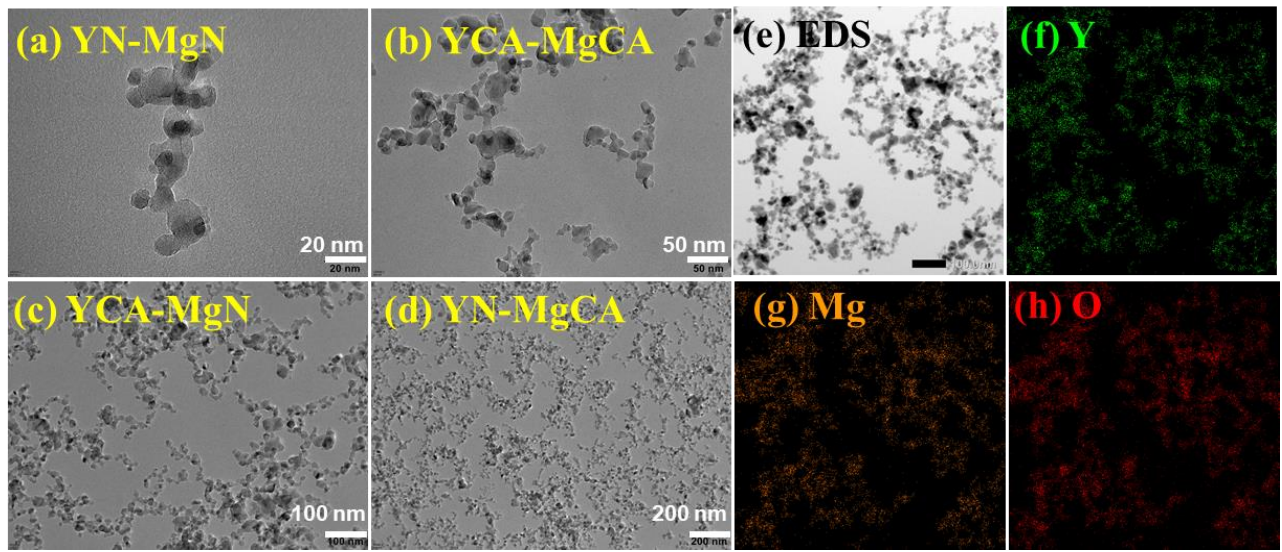

Fig. 4 (a–d) TEM images of particles synthesized from YN-MgN, YCA-MgCA, YCA-MgN, and YN-MgCA precursor combinations, respectively; (e–h) EDS elemental mapping of $Y_2O_3$–MgO particles.

Table 3 Particle sizes derived from XRD Rietveld refinement ($d_{XRD}$) and BET ($d_{BET}$) from different precursor combinations.

|  | $d_{XRD}$ of MgO (nm) | $d_{XRD}$ of C-$Y_2O_3$ (nm) | $d_{XRD}$ of M-$Y_2O_3$ (nm) | $d_{BET}$ of particles (nm) |
|---|---|---|---|---|
| YN-MgN | 12.2 | 23.9 | 7.7 | 20.8 |
| YCA-MgCA | 11.2 | 21.1 | 9.7 | 18.2 |
| YCA-MgN | 11.8 | 23.4 | 9.9 | 20.0 |
| YN-MgCA | 11.9 | 22.4 | 10.1 | 19.2 |

*3.2 Solid solubility of MgO in $Y_2O_3$*

To investigate the solid solubility of MgO in $Y_2O_3$, we designed two flame conditions by varying the dispersion gas (air vs. $O_2$), while maintaining a fixed total flow rate of 15 L/min. The corresponding equivalence ratios were calculated to be 3.56 for air and 0.84 for oxygen, indicating fuel-rich and slightly fuel-lean conditions, respectively. The flame temperature was measured using a type-B thermocouple and corrected for radiative losses. As shown in Fig. 5, the maximum flame temperature reached 1357 °C when using air as the dispersion gas, with an estimated cooling rate of 63 °C/cm. In contrast, when pure oxygen was used, the flame temperature exceeded the upper limit of the thermocouple; the highest measurable temperature was 1907 °C. The corresponding cooling rate was approximately 173 °C/cm, about 2.7 times higher than that under air.

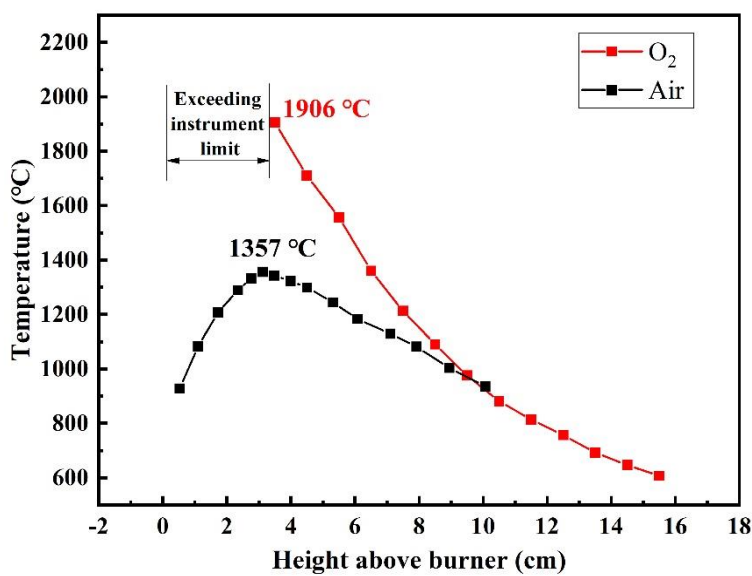

Fig. 5 Flame temperatures under air and O$_2$ as dispersion gases.

Fig. 6 shows the XRD patterns of samples synthesized with varying MgO molar fractions under air dispersion gas. All samples exhibit sharp and well-defined diffraction peaks, indicating good crystallinity. For compositions containing 0–90 mol% MgO, characteristic peaks of cubic Y$_2$O$_3$ are clearly observed at 2θ = 16.7°, 29.2°, 33.8°, and 48.5°. At 100 mol% MgO, only the diffraction peaks corresponding to cubic MgO at 2θ = 42.9° and 62.3° remain. As highlighted in Fig. 6(b), a gradual shift of the (222) peak of C-Y$_2$O$_3$ at 2θ = 29.2° toward higher angles is observed with increasing MgO content, becoming more pronounced above 70 mol%. This shift indicates a decrease in the lattice spacing due to partial incorporation of smaller Mg$^{2+}$ ions (r = 0.72 Å) into the Y$_2$O$_3$ lattice (Y$^{3+}$, r = 0.90 Å), suggesting the formation of a solid solution. In the enlarged regions at 2θ = 41–45° and 61–64°, only C-Y$_2$O$_3$ peaks (2θ = 43.5° and 61.8°) are visible at low MgO contents (0–30 mol%). As the MgO content increases to 40–50 mol%, the characteristic diffraction peaks of MgO at 42.9° and 62.3° start to emerge, which gradually intensify with further MgO addition, while the corresponding $Y_2O_3$ peaks progressively weaken.

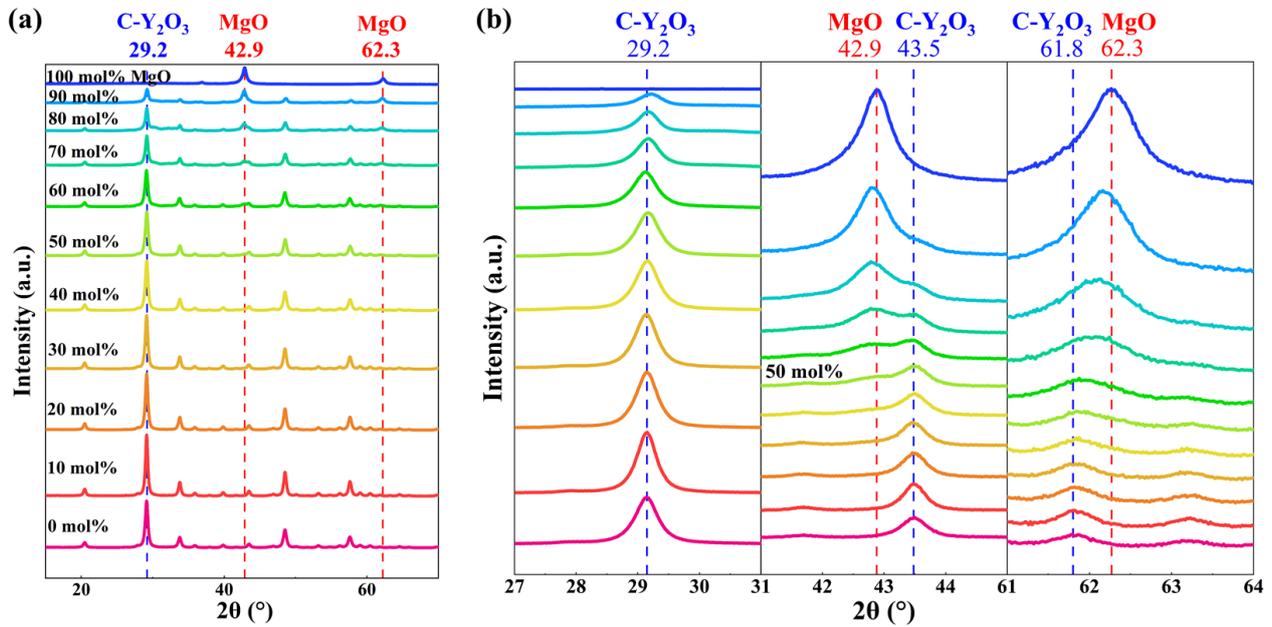

Fig. 6 (a) XRD patterns of samples with MgO molar ratios ranging from 0 to 100% synthesized with air as the dispersion gas; (b) Enlarged views in the 2θ ranges of 27–31°, 41–45°, and 61–64°.

Fig. 7 displays the XRD patterns of samples synthesized under $O_2$ dispersion gas with varying MgO molar fractions. Compared to the samples prepared under air, the diffraction peaks are broadened and less intense, indicating a reduced crystallinity. Moreover, strong peaks corresponding to M-$Y_2O_3$ are observed. As shown in the enlarged region of Fig. 7(b) (27.5–34.5°), the diffraction peaks of C-$Y_2O_3$ at 2θ = 29.2° and 33.8° are marked with blue dashed lines, while those of M-$Y_2O_3$ at 2θ = 28.8°, 29.8°, 30.5°, 31.7°, 32.2°, and 33.1° are highlighted with black dashed lines. With increasing MgO content, the C-$Y_2O_3$ peaks gradually diminish, whereas M-$Y_2O_3$ peaks become increasingly dominant, indicating a phase evolution from cubic- to monoclinic-$Y_2O_3$ as the major phase. In the regions of 2θ = 41–45° and 61–64°, the characteristic peaks of MgO at 42.9° and 62.3° emerge when the MgO content reaches approximately 60 mol%. These peaks exhibit a shift toward lower diffraction angles compared to the pure MgO sample (100 mol%), suggesting an expansion in lattice spacing due to $Y^{3+}$ incorporation into the MgO lattice.

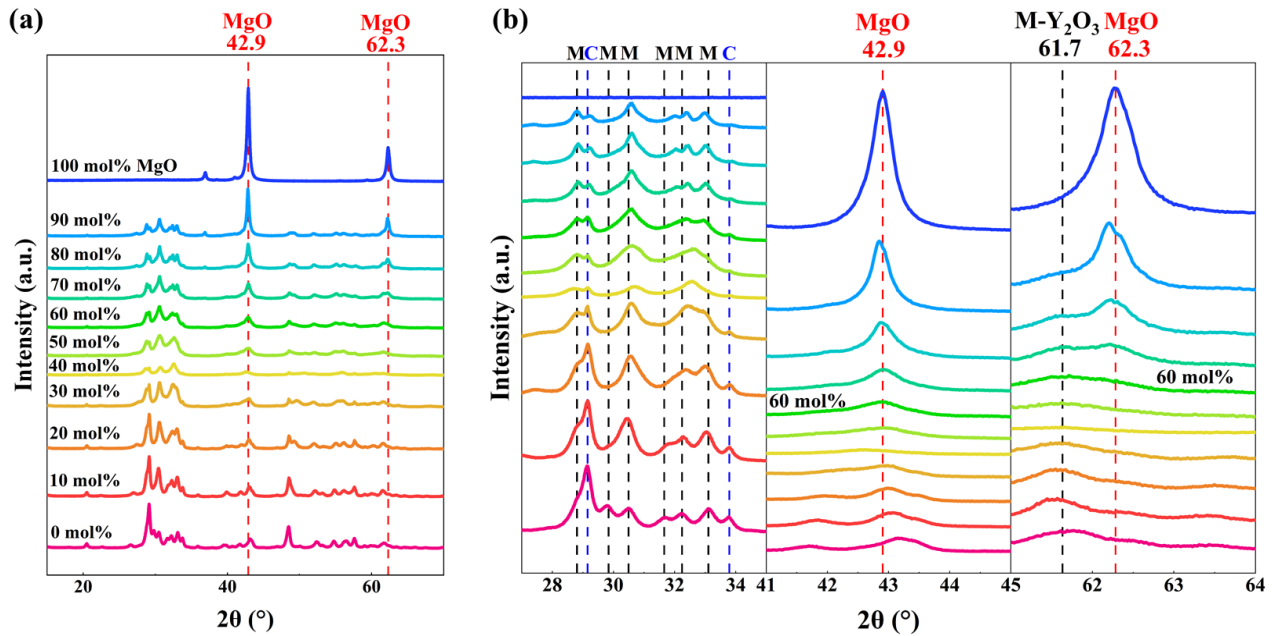

Fig. 7 (a) XRD patterns of samples with MgO molar ratios ranging from 0 to 100% synthesized with $O_2$ as the dispersion gas; (b) Enlarged views in the 2θ ranges of 27.5–34.5°, 41–45°, and 61–64°.

Quantitative analysis of phase compositions under different dispersion gases further elucidates the impact of flame conditions on solid solubility, as shown in Fig. 8. Two key observations can be drawn. First, the fraction of M-$Y_2O_3$ relative to total $Y_2O_3$ increases with MgO content. Under air atmosphere, this ratio rises from 0% to 57.4%, while under $O_2$ it increases from 69.5% to 91.8% across the range of 0-90 mol% MgO. This suggests that $Mg^{2+}$ incorporation destabilizes the $Y_2O_3$ lattice, facilitating the formation of the metastable monoclinic phase. Second, according to Rietveld refinement, no separate MgO phase is detected below 30 mol% MgO under air, or below 50 mol% under $O_2$, indicating complete solid solubility in these ranges. This significantly exceeds the solubility limit reported in the equilibrium phase diagram (Fig. 1), where $Y_2O_3$ and MgO are immiscible at room temperature and exhibit only ~7 mol% solubility at the eutectic point (~2100 °C). Interestingly, although cubic MgO and cubic $Y_2O_3$ share the same lattice type, our results show greater MgO solubility in monoclinic $Y_2O_3$ under $O_2$. This indicates that solubility is governed less by structural compatibility and more by the degree of atomic mixing at high temperature and the rapid quenching intrinsic to flame synthesis, which helps preserve non-equilibrium solid solution structures.

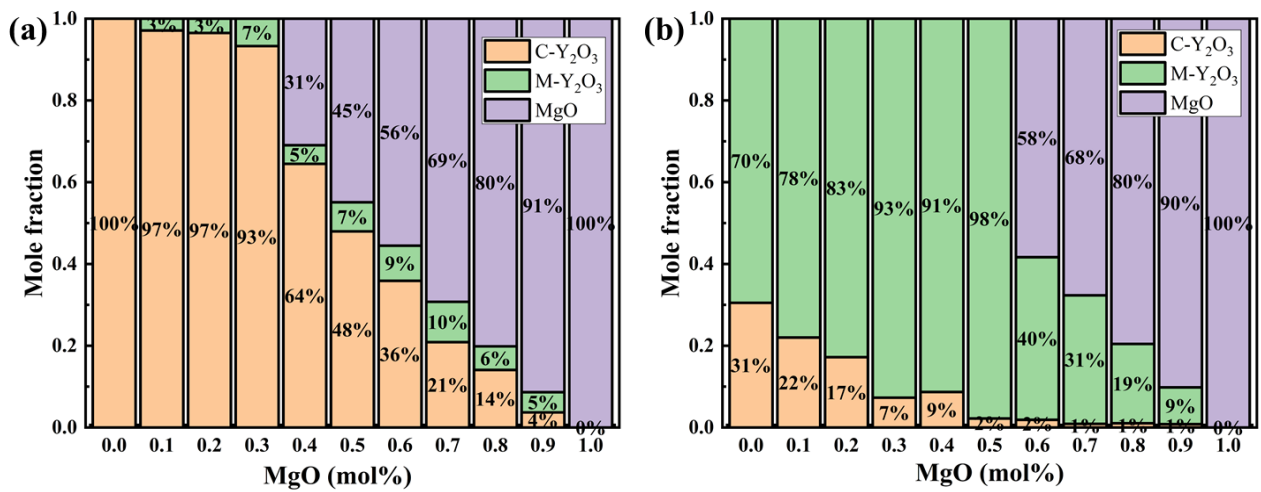

Fig. 8 Phase composition of samples with varying MgO molar ratios from 0 to 100%; (a) synthesized with air as the dispersion gas; (b) synthesized with $O_2$ as the dispersion gas.

The evolution of lattice parameters under air atmosphere is summarized in Table 4. With increasing MgO fraction, the lattice parameter of cubic $Y_2O_3$ gradually decreases from 10.607 Å (0 mol% MgO) to 10.579 Å (90 mol% MgO), consistent with substitutional incorporation of smaller $Mg^{2+}$ ions into the $Y_2O_3$ lattice. Similarly, the lattice parameter of MgO in the presence of $Y_2O_3$ (4.219–4.224 Å) is larger than that of pure MgO (4.203 Å), confirming the expansion due to $Y^{3+}$ doping. Additionally, the unit cell volume of monoclinic $Y_2O_3$ notably decreases from 460.3 Å³ at 10 mol% MgO to approximately 411 Å³ at higher MgO content (≥70 mol%), reflecting the structural distortion associated with increasing Mg incorporation.

Table 4 Lattice parameters of different phases with varying MgO molar ratios under air

| MgO fraction | Lattice Parameter of C-$Y_2O_3$ (Å) | Unit Cell Volume of M-$Y_2O_3$ (Å$^3$) | Lattice Parameter of MgO (Å) | R |
|---|---|---|---|---|
| 0 | 10.607 | / | / | 4.4% |
| 0.1 | 10.603 | 460.3 | / | 3.5% |
| 0.2 | 10.605 | 460.1 | / | 3.4% |
| 0.3 | 10.604 | 449.6 | / | 3.7% |
| 0.4 | 10.604 | 441.2 | 4.224 | 3.3% |
| 0.5 | 10.602 | 434.3 | 4.219 | 3.2% |
| 0.6 | 10.605 | 412.9 | 4.222 | 3.6% |
| 0.7 | 10.600 | 411.3 | 4.222 | 3.7% |
| 0.8 | 10.596 | 411 | 4.223 | 3.7% |
| 0.9 | 10.579 | 411.2 | 4.220 | 4.0% |
| 1 | / | / | 4.203 | 5.3% |

The effect of MgO molar ratio on particle morphology is shown in Fig. 9 (air atmosphere) and Fig. S1 ($O_2$ atmosphere). Since the particle morphology synthesized under air and oxygen conditions is similar, taking the example of air atmosphere in Fig. 9. All samples exhibit nanoscale, irregularly shaped particles with sintering necks, indicating solid, crystalline structures. As the MgO content increases, the primary particle size remains relatively consistent, with no signs of hollow morphology. At higher MgO contents, flake-like structures appear, as highlighted by the red dashed boxes in Fig. 9(j, k). To clarify the composition of these flake-like structures, EDS mapping was performed, as shown in Fig. 10. The results reveals that the flake-rich regions (highlighted by the yellow box) are dominated by Mg signals with negligible Y content, confirming the formation of Mg-rich, Y-free domains. Combined with their instability under electron beam irradiation, these structures are attributed to magnesium hydroxide ($Mg(OH)_2$) formed due to the hydration of unincorporated excess MgO.

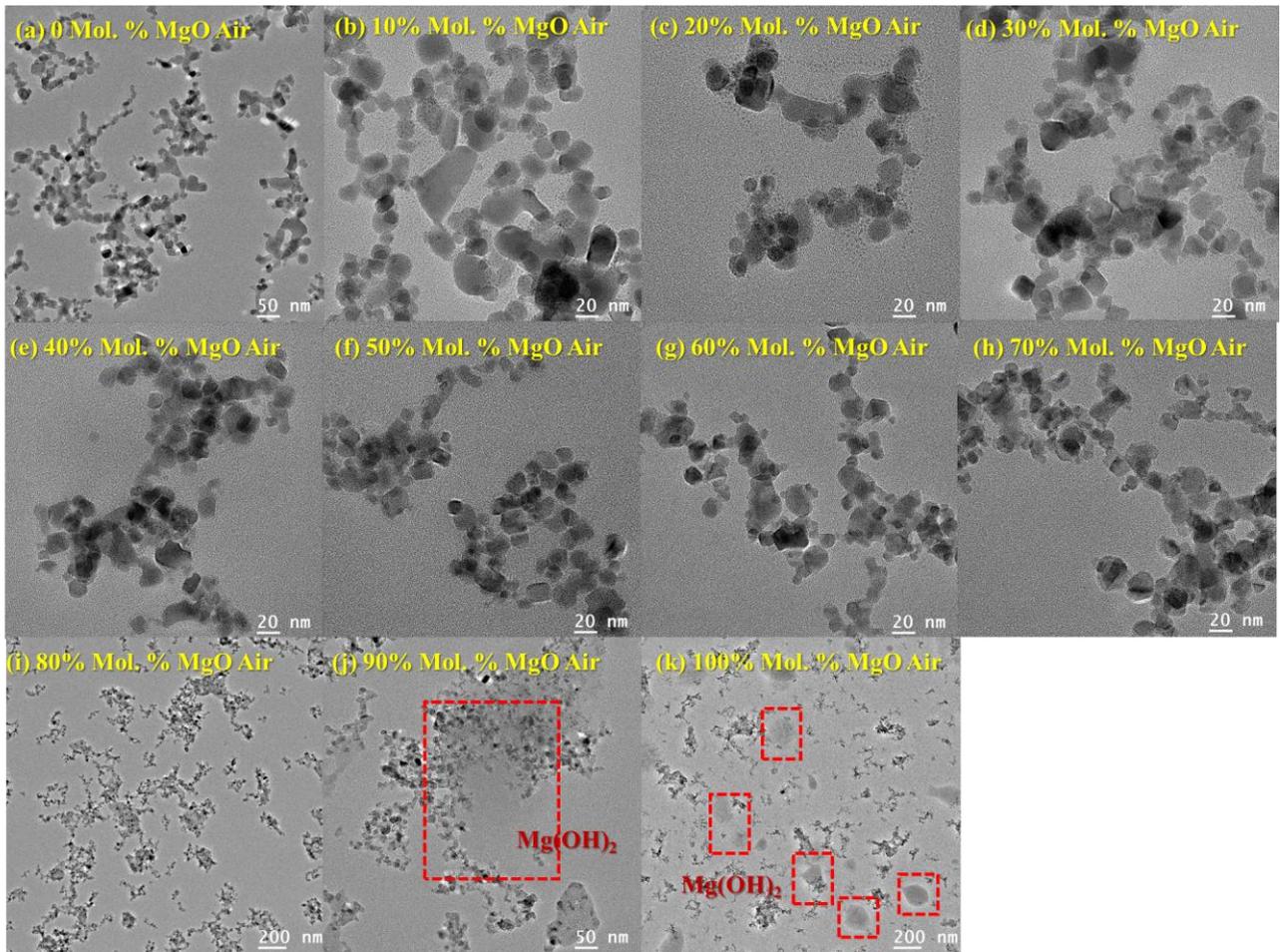

Fig. 9 Morphology of flame-synthesized particles with MgO molar fractions ranging from 0% to 100% under air as the dispersion gas

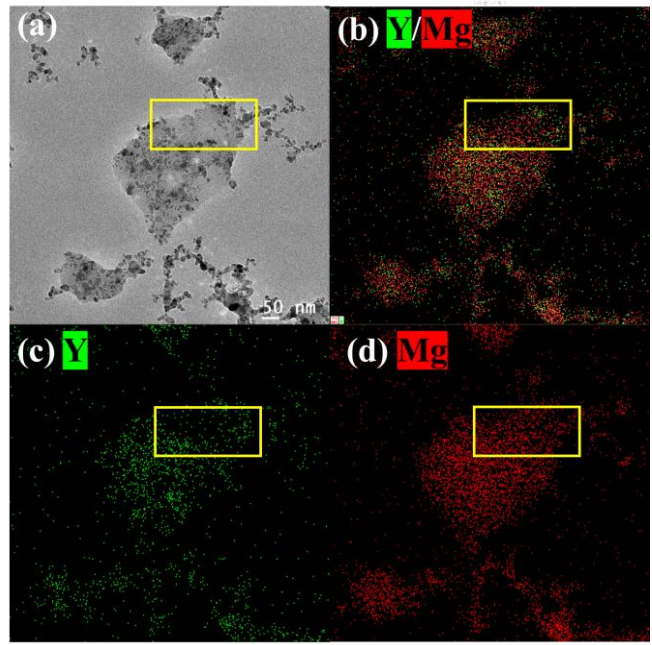

Fig. 10 Elemental EDS mappings of flake-like structures formed in the samples

Fig. 11 summarizes the variation in particle size with increasing MgO molar ratio. From the line plots of BET-derived particle sizes, it is observed that particles synthesized under oxygen atmosphere (red line) exhibit slightly smaller sizes compared to those produced under air (black line). The XRD-derived crystallite size of C-$Y_2O_3$ remains relatively stable in the range of 20–25 nm for MgO fractions up to 90 mol%, while a notable drop is observed in the M-$Y_2O_3$ crystallite size, decreasing from ~24 nm at 10 mol% to below 5 nm at ≥40 mol%, indicating a suppressed particle growth or finer nucleation. The crystallite size of MgO remains nearly constant (~10–13 nm), independent of composition.

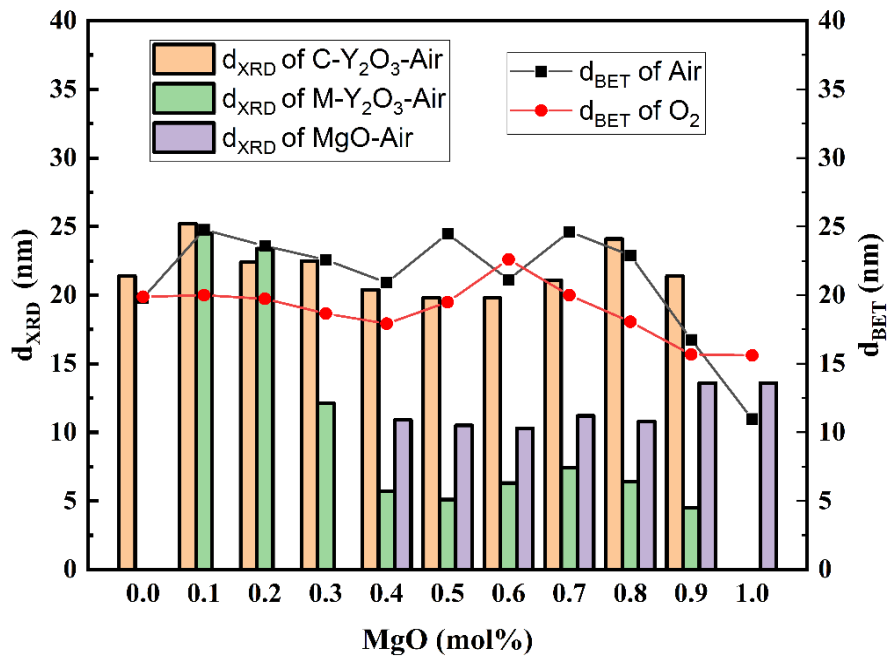

Fig. 11 Variation of particle size as a function of MgO molar fraction

*3.3 Sintering of composite ceramic*

*3.3.1 Effect of powder deagglomeration*

Powder deagglomeration plays a critical role in achieving high-density and high-transparency ceramics. The agglomerates may act as rigid clusters or even form arch-like structures, inhibiting uniform shrinkage and leaving residual porosity in the sintered body[34]. Flame-synthesized powders tend to form soft agglomerates due to their nanoscale size and high surface energy. Therefore, two dispersion techniques—ball milling and high-shear homogenization—were compared with untreated powders.

The influence of deagglomeration on optical and microstructural properties is summarized in Fig. 12. The transmittance spectra in Fig. 12 (a) show that without HIP (dashed lines), powders with no deagglomeration treatment exhibited negligible transmittance across the 3–5 μm range, despite a relative density of 98.3% (Fig. 12(b)), indicating that residual porosity or inhomogeneous microstructure strongly scatters mid-IR light. Homogenization improved both density (99.0%) and $T_{3-5\mu m}$ to 29%, while ball-milled powders showed the best performance, achieving 99.3% density and $T_{3-5\mu m}$ of 48%. After HIP treatment, further densification occurred across all samples (solid lines), particularly enhancing the transparency. Ball milling again proved most effective, yielding a peak transmittance ($T_{MAX}$) of 83.3% and $T_{3-5\mu m}$ up to 79.5%.

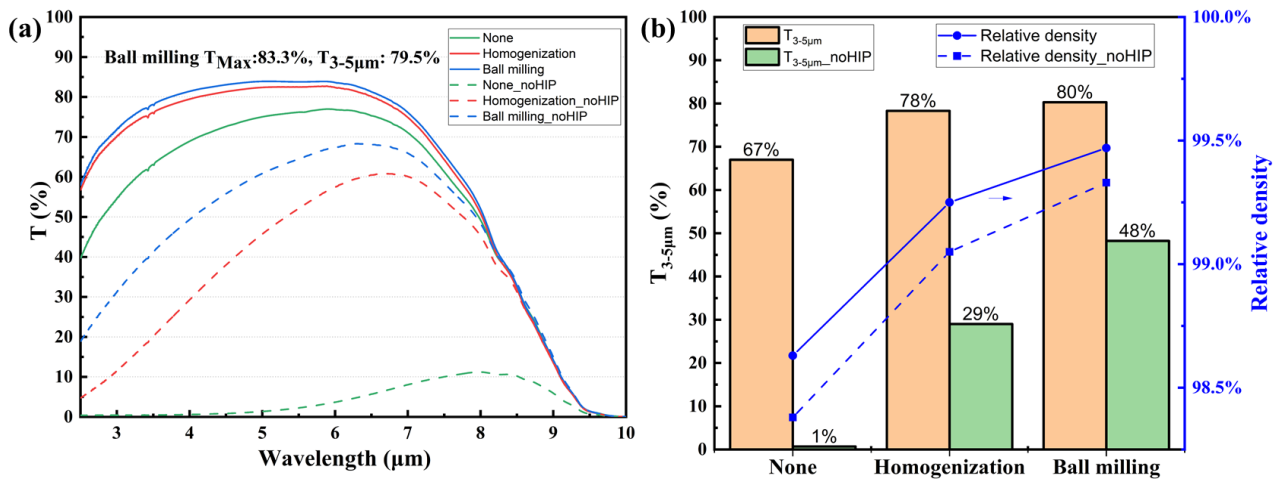

Fig. 12 (a) IR transmittance spectra of $Y_2O_3$–MgO ceramics fabricated from powders without deagglomeration (None), homogenized, and ball-milled. (b) Comparison of average transmittance in the 3–5 μm ($T_{3-5\ \mu m}$) and relative density (right axis) for different deagglomeration methods.

These optical results are further supported by the microstructural observations shown in Fig. 13. All samples appear fully dense, but clear differences in phase distribution are observed. In both the ball-milled (Fig. 13(a)) and homogenized (Fig. 13(b)) ceramics, the two phases—white $Y_2O_3$ and black MgO—are uniformly distributed, indicating good mixing before sintering. In contrast, the ceramic prepared without deagglomeration (Fig. 13(c)) exhibits enrichment of a single phase, as highlighted by the red circle. Additionally, intragranular pores are observed within $Y_2O_3$ grains (green arrows), which are likely formed during sintering as a result of arch-like structures originating from agglomerated particles, which trap voids that are difficult to eliminate. These results confirm that adequate powder deagglomeration is essential for achieving both high densification and optical quality, with ball milling being the most effective method among those tested, likely due to its stronger mechanical dispersion and longer processing time, resulting in more thorough breakdown of agglomerates.

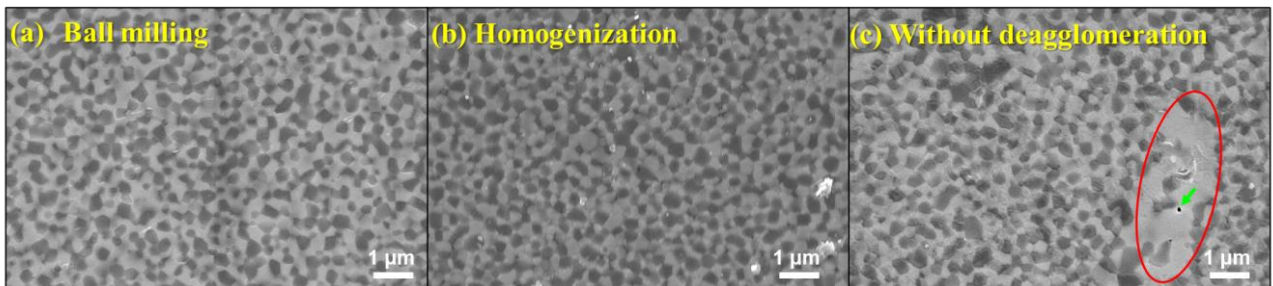

Fig. 13 SEM images of $Y_2O_3$–MgO ceramics fabricated from powders (a) ball-milled, (b) homogenized, and (c) without deagglomeration.

3.3.2 Effect of vacuum sintering temperature

The infrared transmittance and microstructural evolution of vacuum-sintered samples at different temperatures are summarized in Fig. 14 and Fig. 15. All samples in this section were sintered without subsequent HIP treatment. As shown in Fig. 14(a), the infrared transmittance increases from 1250 °C to 1300 °C, reaching a max average transmittance of 55% in the 3–5 μm range at 1300 °C (Fig. 14(b)). This improvement correlates with a sharp rise in relative density, from 96.9% at 1250 °C to 99.0% at 1300 °C, indicating enhanced densification. However, further increasing the sintering temperature to 1350 °C and 1400 °C leads to a slight decrease in transmittance, despite a continued increase in relative density. This decoupling suggests that densification is no longer the limiting factor for optical performance; instead, excessive grain growth and microstructural coarsening may introduce light-scattering defects.

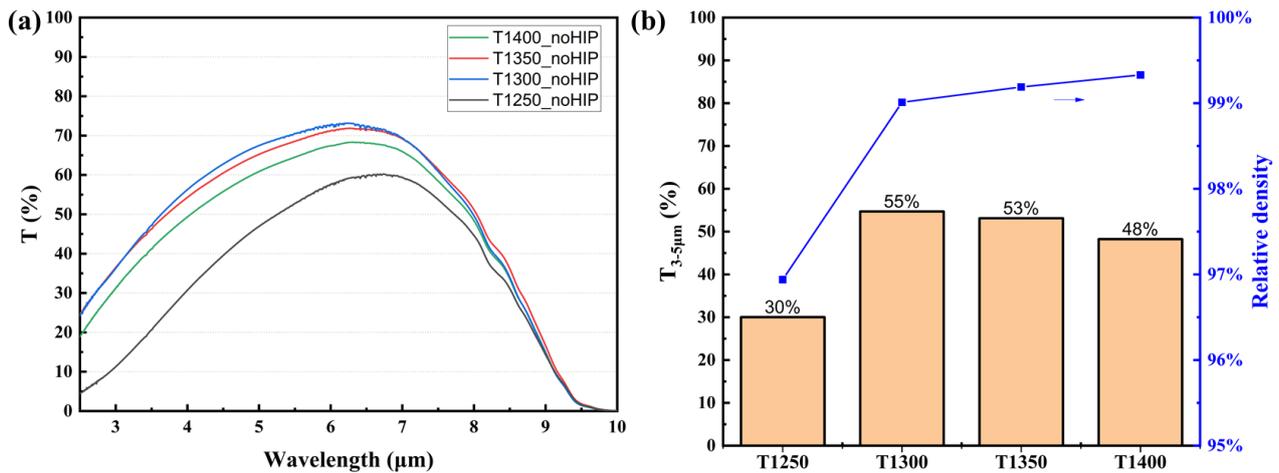

Fig. 14 (a) IR transmittance spectra of $Y_2O_3$–MgO ceramics vacuum-sintered at 1250–1400 °C for 5 h without HIP treatment. (b) Comparison of 3–5 μm average transmittance and relative density as a function of sintering temperature.

The SEM images in Fig. 15(a–d) provide direct evidence of these trends. At 1250 °C, numerous residual pores are observed (highlighted by red circles), which accounts for the poor densification and low transmittance. Upon increasing the sintering temperature to 1300 °C, the pores are effectively eliminated, and exhibits a highly dense and uniform two-phase microstructure (brighter $Y_2O_3$ grains and darker MgO grains), suggesting a good pinning effect. At 1400 °C, although visible porosity is absent, the microstructure becomes coarser and more inhomogeneous. The blue boxes in Fig. 15(d) highlight regions where $Y_2O_3$ grains are interconnected and coarsened, indicating that the Zener pinning effect has partially failed, allowing abnormal grain growth. This interpretation is further supported by the grain size distributions shown in Fig. 15(e). The average grain size increases modestly from 137 ± 54 nm at 1250 °C to 166 ± 62 nm at 1300 °C, with a relatively narrow distribution. A notable coarsening occurs at 1350 °C, where the average grain size rises significantly to 208 ± 81 nm, along with a broadening in distribution. At 1400 °C, this trend continues, reaching 251 ± 95 nm, indicating accelerated grain growth and increasingly inhomogeneous microstructure, which result in increased scattering and therefore reduced transmittance.

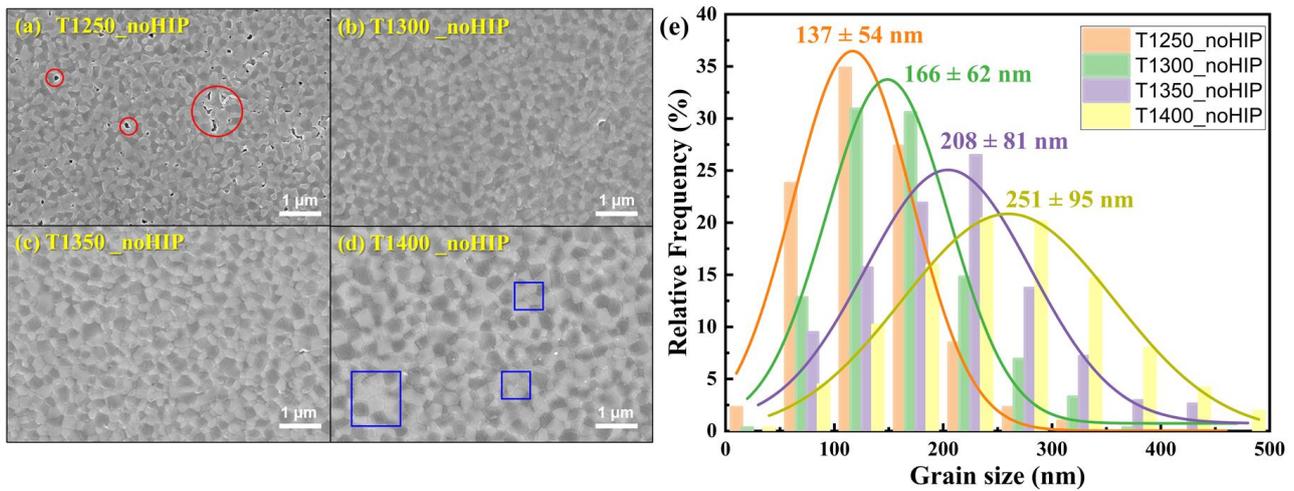

Fig. 15 (a–d) SEM images of polished surfaces for samples sintered at 1250–1400 °C (without HIP); (e) Grain size distribution obtained from SEM images.

3.3.3 Effect of HIP temperature

To further enhance optical performance and suppress residual porosity, hot isostatic pressing (HIP) was applied after vacuum sintering. Based on the previous section, two vacuum pre-sintering temperatures (1250 and 1300 °C) were selected to investigate the effect of HIP temperature (1200–1300 °C) on microstructure and transmittance.

As shown in Fig. 16(a), for both vacuum sintering temperatures, increasing HIP temperature leads to improved IR transmittance. The best optical performance was obtained for the sample treated with 1300 °C vacuum sintering followed by 1300 °C HIP, with $T_{MAX}$ of 84.6% at 5 μm and $T_{3-5\mu m}$ of 82.3%. This value exceeds those reported in the literature for ceramics of same thickness, including: an average transmittance of 71% in the 3–7 μm range for ball-milled powders processed by hot pressing[35]; a maximum transmittance close to 80% in the 2.5–6 μm range for colloidally synthesized powders sintered by SPS[27]; $T_{MAX}$ =80% at 5 μm for sol–gel synthesized powders processed by two-step SPS[36]; $T_{MAX}$ =82% in the 3–6 μm range for sol–gel powders sintered by hot pressing[37]; $T_{MAX}$ =78% at 6 μm for SHS powders densified via microwave sintering[20]; and $T_{MAX}$ =80.9% at 5 μm for SHS powders sintered by SPS[10].

Notably, at lower HIP temperatures (e.g., 1200–1250 °C), samples pre-sintered at 1250 °C show better transmittance than those pre-sintered at 1300 °C. This is likely because higher pre-sintering temperatures result in near-full densification (e.g., 98.8% at 1300 °C), leaving insufficient driving force for further densification during HIP. Consequently, HIP mainly promotes grain growth rather than pore elimination, compromising optical quality. These results highlight the importance of balancing vacuum and HIP temperatures to avoid over-sintering and ensure optimal transparency.

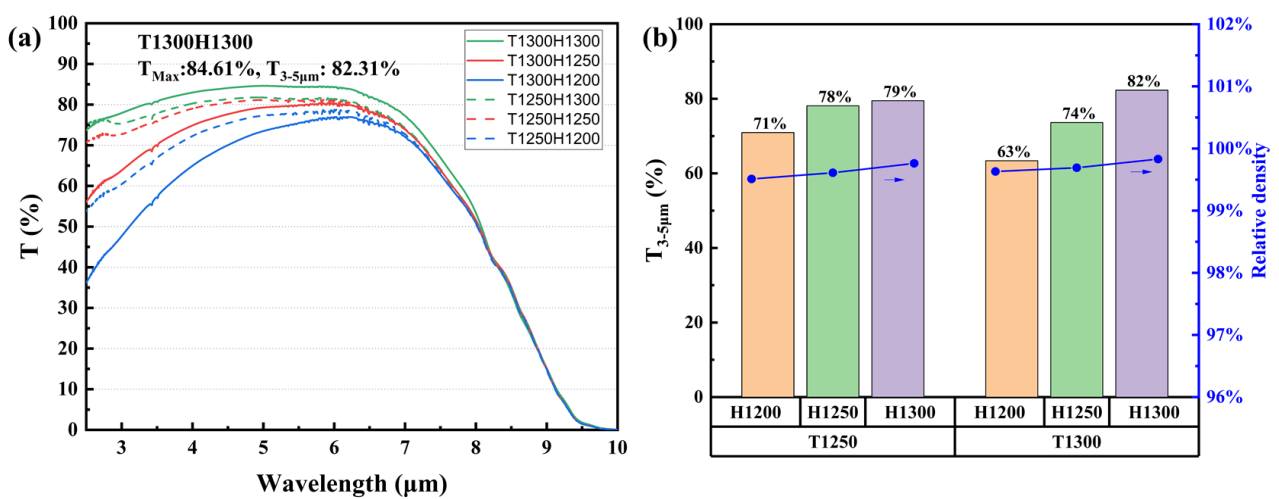

Fig. 16 (a) IR transmittance spectra of Y$_2$O$_3$–MgO ceramics prepared at two vacuum sintering temperatures (1250 and 1300 °C), followed by HIP at 1200–1300 °C. (b) Comparison of 3–5 μm average transmittance and relative density under different HIP temperatures.

To identify the main factors contributing to transmittance loss, the theoretical transmittance was calculated and compared with the experimental results, by using Eq.(1):

$$T_{Predicted} = \frac{(1-R)^2 e^{-(u_\alpha+u_s)b}}{1-R^2 e^{-2(u_\alpha+u_s)b}} \quad (1)$$

where $R$ is the reflectance at the interface, $R = (1-n)^2/(1+n)^2$. $n$ is the refractive index, $u_\alpha$ is the absorption coefficient, $u_s$ is the scattering coefficient and $b$ is the sample thickness. The values of $n$ and $u_\alpha$ are obtained from the reference[38].

Assuming a thickness of 1 mm, and neglecting scattering, the predicted transmittance without scatter $T_{MAX}$ reaches 86.6% at 5 μm, and $T_{3-5\mu m}$ is 85.7%. Given the high relative density of 99.83%, residual porosity is expected to be minimal. To evaluate whether residual pores could still contribute to scattering, Mie theory was applied to estimate the scattering loss under the assumption of independent scattering[39]. The scattering coefficient $u_s$ is given by:

$$u_s = \frac{3\varphi Q_{sca}}{2D} \quad (2)$$

where φ is the volume fraction, $Q_{sa}$ is the scattering efficiency calculated by MIE theory, and D is the pore diameter. As no large voids (>100 nm) were observed in SEM images, simulations were performed assuming a porosity of 0.17% and pore diameters of 25, 50, and 75 nm. As shown in Table 5, the resulting scattering losses were marginal, confirming that the deviation from predicted transmittance is primarily attributable to grain boundary-related scattering, and potential contributions from enhanced absorption associated with defects in lattice.

Table 5 Comparison between predicted and measured transmittance and estimated pore scattering losses for sample thickness of 1mm.

| | Predicted transmittance | Measured transmittance | ΔT ($T_{Predicted}$ − | Pore scattering | Pore scattering | Pore scattering |
| --- | --- | --- | --- | --- | --- | --- |

|  | without scatter | of T1300H1300 | $T_{\text{Measured}}$ | loss (D=25 nm) | loss (D=50 nm) | loss (D=75 nm) |
|---|---|---|---|---|---|---|
| $T_{5\mu m}$ | 86.6% | 84.6% | 2% | 0.001% | 0.01% | 0.03% |
| $T_{3-5\mu m}$ | 85.7% | 82.3% | 3.4% | 0.004% | 0.03% | 0.1% |

The microstructural evolution under different HIP temperatures is shown in Fig. 17(a–f). As discussed previously, HIP at 1200 and 1250 °C is insufficient to fully eliminate residual pores. In both the T1250 and T1300 series, clearly visible voids are present in the H1200 and H1250 samples, as marked by red circles. In contrast, HIP at 1300 °C leads to fully dense structures for both pre-sintering conditions, with no visible porosity observed in Fig. 17 (c, f). Across all samples, the two phases are distributed homogeneously. The evolution in grain size distribution is further quantified in Fig. 18(a–b). For both T1250 and T1300 groups, the average grain size increases with HIP temperature. The T1250 series maintains relatively narrow distributions, with grain sizes ranging from 143 nm (HIP at 1200 °C) to 178 nm (HIP at 1300 °C). In contrast, the T1300 group exhibits more pronounced grain growth and broader size distributions, reaching up to 213 nm at the highest HIP temperature. These results confirm that while high HIP temperature is essential for eliminating porosity. When pre-sintering is already nearly full-density, highly densified during pre-sintering, HIP provides limited densification potential, and the process becomes dominated by grain coarsening rather than pore removal, causing a decrease in transmission. Therefore, the microstructure should be carefully engineered through a balanced combination of vacuum sintering and HIP parameters to achieve both full densification and fine grain structures.

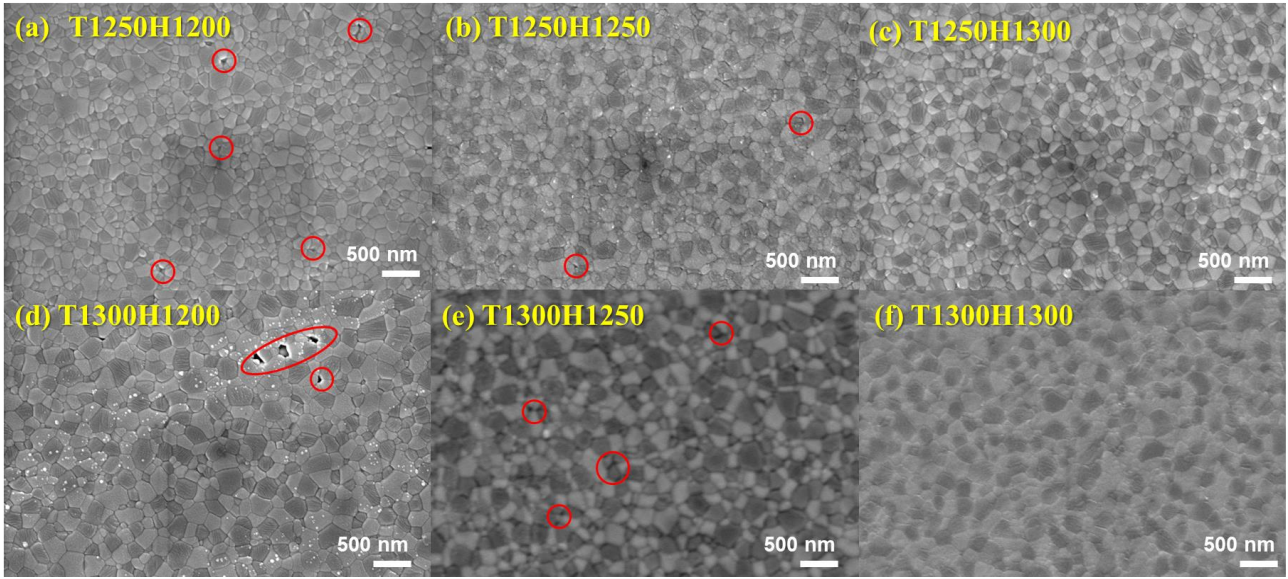

Fig. 17 (a–f) SEM images of $Y_2O_3$–MgO ceramics prepared at two vacuum sintering temperatures (1250 and 1300 °C), followed by HIP at 1200–1300 °C

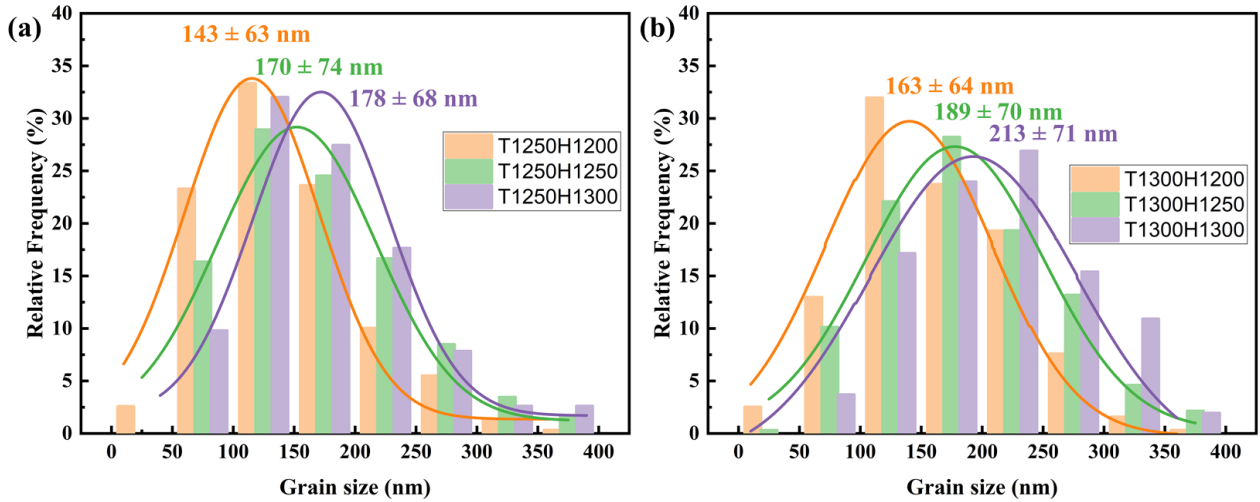

Fig. 18 Grain size distribution of $Y_2O_3$–MgO ceramics after HIP at different temperatures: (a) samples with 1250 °C vacuum sintering; (b) samples with 1300 °C vacuum sintering.

3.3.4 Effect of initial powder characteristics

To investigate the influence of initial powder characteristics on the final microstructure and optical properties, three types of $Y_2O_3$–MgO nanopowders with distinct phase compositions were compared. The first powder (labeled as 71% Cubic) was synthesized from YN-MgN precursors under air atmosphere and consisted predominantly of cubic $Y_2O_3$ (71%), with a BET particle size of 20.8 nm. The second powder (95% Monoclinic) was produced using the same precursors but under $O_2$ atmosphere, resulting in a monoclinic-rich phase (95%) and a smaller BET particle size of 18.1 nm. The third powder (100% Cubic-Annealing) was obtained by annealing the monoclinic-rich powder at 1200 °C for 1 h in air, which fully transformed it into the cubic phase while increasing the particle size to 62.4 nm. The XRD patterns of the three powders are shown in Fig. S2.

Fig. 19 presents the transmittance spectra and relative densities of ceramics prepared from powders with different initial powders. Among the three samples, the ceramic fabricated from 100% Cubic-Annealing powder exhibits the highest 3–5 μm transmittance. It is noteworthy that the 95% Monoclinic sample shows a slower drop in transmittance in the short-wave IR region (2.5–3.0 μm) and superior performance in the near-infrared band. As highlighted in the inset of Fig. 19(a), the transmittance at 1064 and 1550 nm reaches 26.2% and 56.2%, respectively. Further insights can be drawn from the relative density data in Fig. 19(b). Even without HIP treatment, the 95% Monoclinic sample attains a relative density of 99.5%, the highest among all compositions, indicating excellent sintering activity. This enhanced sinterability is likely associated with the finer initial particle size and the high solubility of MgO in the $Y_2O_3$ when synthesized in an oxygen atmosphere. Specifically, up to 50 mol% MgO can be fully dissolved, in contrast to only ~30 mol% under air atmosphere as in the 71% Cubic sample. The larger solid solubility range is expected to introduce greater lattice distortion, which increases atomic mobility and promotes densification during sintering. While both the 95% Monoclinic and 71% Cubic samples have comparable primary particle sizes, the latter shows reduced green density and poorer optical performance, underscoring the role of solid-solution in driving sintering kinetics.

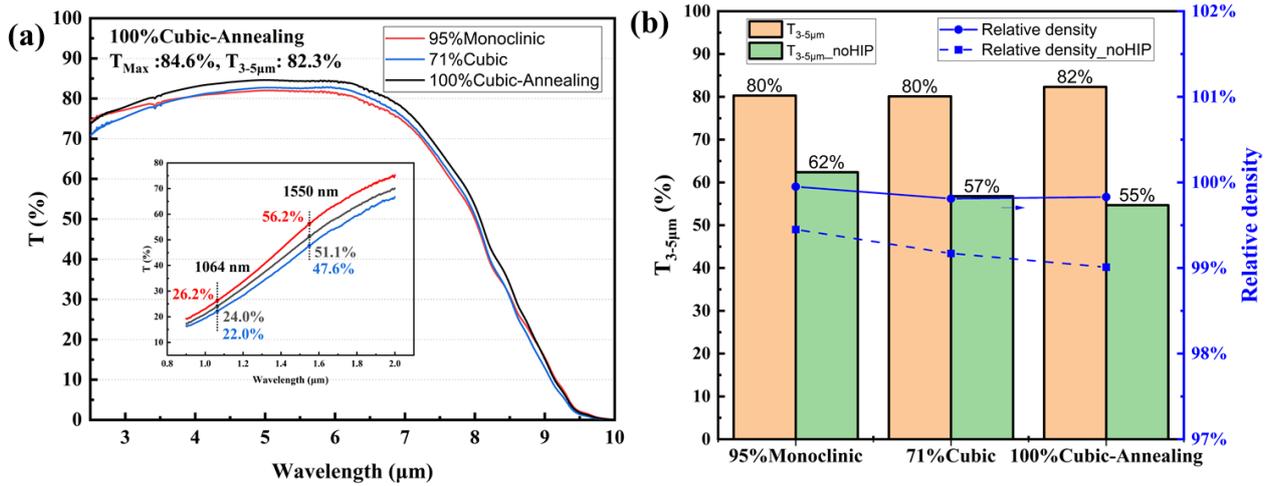

Fig. 19 (a) IR transmittance spectra of ceramics prepared from different initial powders; (b) comparison of 3–5 μm average transmittance and relative density (before and after HIP). The inset in (a) shows the transmittance in the 0.9–2 μm range.

Fig. 20 provides further microstructural and macroscopic comparison. All samples exhibit fully dense structures with no visible residual pores, and the Y$_2$O$_3$ and MgO phases are uniformly distributed. The 95% Monoclinic and 71% Cubic samples exhibit finer grain sizes (~180 nm) than the 100% Cubic-Annealing sample (~213 nm), consistent with their smaller initial particle sizes. However, in Fig. 20(d), significant differences in ceramic appearance can be observed between the three samples. The 95% Monoclinic ceramics exhibit severe cracking and edge breakage, while the 71% Cubic ceramics show moderate cracking, primarily at the edges. In contrast, the 100% Cubic-Annealing ceramics remain intact with no observable defects. The observed damage in 95% Monoclinic and 71% Cubic ceramics samples can be attributed to phase transformation-induced volume changes during sintering. Additionally, the ultrafine particle size in these two powders necessitates a more controlled heating rate to avoid thermal stress accumulation, which was not fully optimized here.

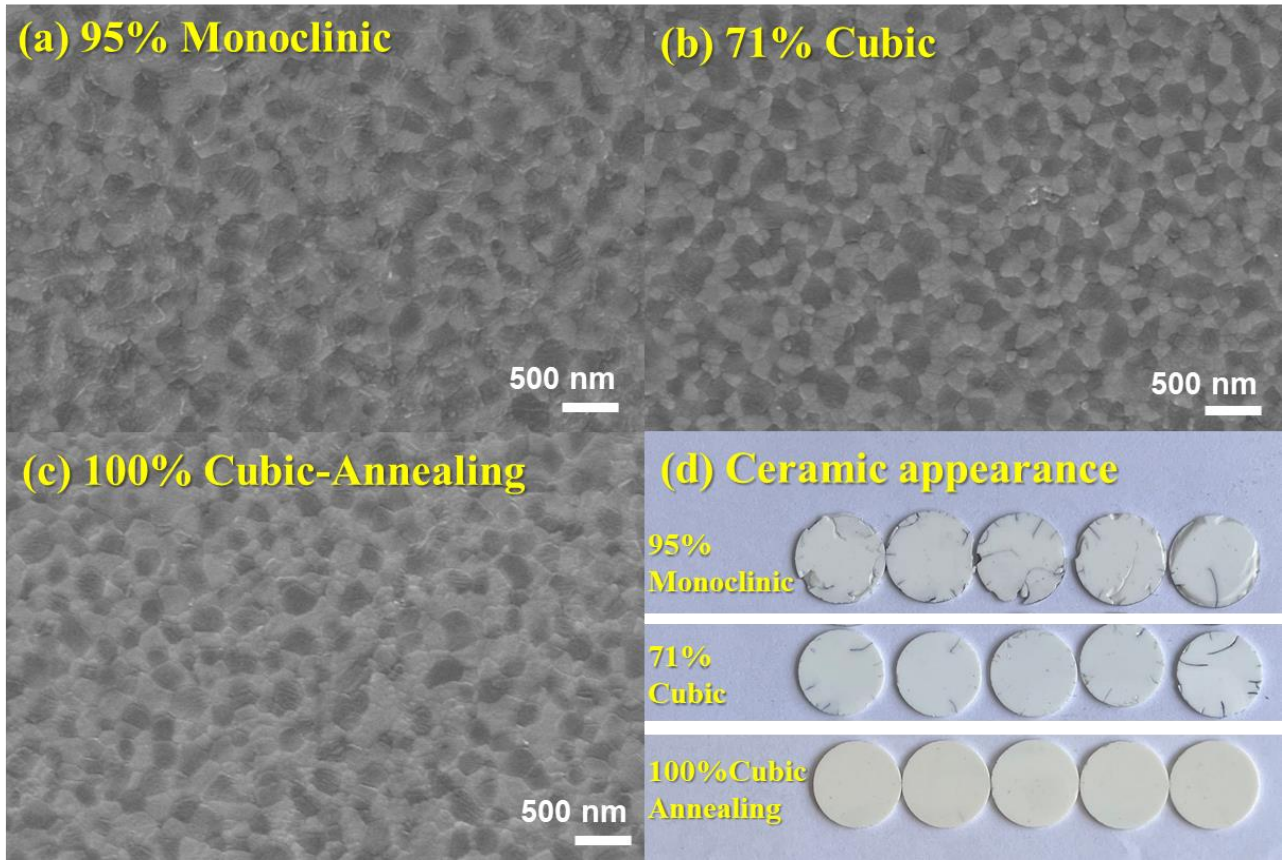

Fig. 20 (a–c) SEM images of ceramics prepared from different initial powder; (d) Ceramic appearance of the corresponding ceramics after polishing

## 4. Conclusion

Through a systematic investigation of precursor chemistry, flame synthesis conditions, and subsequent sintering conditions, we have clarified the critical factors influencing the microstructure and optical performance of $Y_2O_3$–MgO transparent ceramics. Precursor choice significantly affects particle crystal phase, with nitrates favoring cubic $Y_2O_3$ formation and acetates promoting the metastable monoclinic phase. Flame synthesis in oxygen atmosphere provided high temperatures and rapid quenching rates, enabling unprecedented MgO solubility (up to 50 mol%), substantially surpassing equilibrium limits, which enhances atomic mixing and densification kinetics. Powder deagglomeration was essential to achieve high density and optical quality, with ball milling providing optimal results. The combination of vacuum sintering and HIP at 1300 °C yielded the highest optical performance, achieving a maximum transmittance of 84.6% and an average transmittance of 82.3% in the 3–5 μm range. Importantly, excessive pre-sintering reduced the densification driving force during HIP, causing undesirable grain coarsening. Initial powder crystal phases profoundly impacted ceramic performance: monoclinic-dominated powders offered superior sintering activity and near-infrared transparency but suffered from phase-transformation-induced cracking, while annealed cubic-phase powders provided excellent mid-infrared transparency and intact appearance. Overall, this study highlights the necessity of carefully coordinating flame synthesis conditions, initial powder selection, and sintering strategies to produce optimal infrared-transparent $Y_2O_3$–MgO ceramics.

## CRediT authorship contribution statement

**Shuting Lei:** Investigation, Data curation, Methodology, Formal analysis, Writing - original draft. **Yiyang Zhang:** Conceptualization, Writing - review & editing, Funding acquisition, Supervision. **Xing Jin:** Methodology, Resources. **Yanan Li:** Investigation, Validation. **Zhu Fang:** Investigation, Writing - review & editing. **Shuiqing Li:** Resources, Supervision, Funding acquisition, Project administration, Writing - review & editing.

## Declaration of competing interest

The authors declare that they have no known competing financial interests or personal relationships that could have appeared to influence the work reported in this paper.

## Data availability

Data will be made available on request.

# Acknowledgments

This work was supported by the National Natural Science Foundation of China (Grant nos. 52130606, 52322608) and the Space Application System of China Manned Space.